\documentclass[authoryear,12pt]{elsarticle}

\usepackage{amssymb}

\def\astrobj#1{#1}

\newcommand{\mdot}{\ensuremath{\dot{m}\ }}

\newcommand{\hei}{\ensuremath{\mathrm{He~I}}}
\newcommand{\heii}{\ensuremath{\mathrm{He~II}}}
\newcommand{\oiii}{\ensuremath{\mathrm{[O~III]}}}
\newcommand{\oii}{\ensuremath{\mathrm{[O~II]}}}
\newcommand{\oi}{\ensuremath{\mathrm{[O~I]}}}
\newcommand{\sii}{\ensuremath{\mathrm{[S~II]}}}

\newcommand{\feiii}{\ensuremath{\mathrm{[Fe~III]}}}

\newcommand{\nii}{\ensuremath{\mathrm{[N~II]}}}

\newcommand{\fevii}{\ensuremath{\mathrm{Fe~VII}}}

\newcommand{\ergl}{\ensuremath{\mathrm{erg\,s^{-1}}}}
\newcommand{\kms}{\ensuremath{\mathrm{km\,s^{-1}}}}

\newcommand{\Myr}{\ensuremath{\mathrm{Myr}}}
\newcommand{\yr}{\ensuremath{\mathrm{yr}}}
\newcommand{\pc}{\ensuremath{\mathrm{pc}}}
\newcommand{\kpc}{\ensuremath{\mathrm{kpc}}}
\newcommand{\cmc}{\ensuremath{\mathrm{cm^{-3}}}}
\newcommand{\keV}{\ensuremath{\mathrm{keV}}}

\newcommand{\Msun}{\ensuremath{\mathrm{M_\odot}}}

\journal{New Astronomy}

\begin{document}

\begin{frontmatter}



\title{Understanding ULX Nebulae in the Framework of Supercritical Accretion}


\author{P. Abolmasov}

\address{Sternberg Astronomical Institute, Moscow State University
  Moscow, Russia 119992; Email: \emph{pavel\_abolmasov@yahoo.co.uk}}

\begin{abstract}
For a long time, the well-known supercritically accreting binary 
SS433 is being proposed as a prototype for a class of
hypothetical bright X-ray sources that may be identified with the
so-called Ultraluminous X-ray sources (ULXs) in nearby galaxies
 or at least with part of them.
Like SS433, these objects should be associated with optical nebulae,
powered by both radiation of the central source and its wind or jet
activity. Indeed, around many ULXs, bright optical nebulae (ULX
Nebulae, ULXNe) are found. 
Here, we use SS433 as a prototype for the power source creating the
nebulae around ULXs.
Though many factors are important such as the structure of the host
star-forming region and the possible supernova remnant formed together
with the accreting compact object, we show that most of the
properties of ULXNe may be explained by an
SS433-like system evolving for up to about one million years 
in a constant density environment.
The basic stages of evolution of a ULX Nebula include a
non-spherical HII-region with a central cavity created by non-radiative
shock waves, an elongated or bipolar shock-powered nebula created by jet activity
and a large-scale quasi-spherical bubble. 
\end{abstract}


\end{frontmatter}


\section{Introduction}
\label{sec:intro}

The issue of supercritical accretion is more than three
decades old. It was first considered in the seminal paper by
\citet{SS73} who pointed out the significance of Eddington limit for
accretion processes. A mass accretion rate exceeding the critical, or
Eddington, value leads to Eddington luminosity limit violation in the
framework of standard disc accretion. The disc may become either
advective \citep{abram80} or outflow-dominated (see \citet{poutanen}
and references therein) if the critical accretion rate is exceeded.

Supercritical accretors are expected to
appear among massive black-hole X-ray binaries during thermal and
sometimes nuclear-timescale mass transfer
\citep{RPP}. The released power is in excess of $\sim 10^{39}\ergl$
(Eddington luminosity for a conventional 10\Msun\ black hole), that makes
supercritical accretors in nearby galaxies prospective targets for
X-ray, UV and optical observations. 
In different models, however, the exact observational properties
differ. Both outflows and advection (advective discs are geometrically
thick) taken apart make sources anisotropic and
soften their spectral energy distributions (SED), thereby increasing
the contribution of extreme ultraviolet radiation in the overall energy
output. That makes supercritical accretors possible energy sources for
bright optical nebulae \citep{katz86,AKK}.

\astrobj{SS433} is known as a unique Galactic object in the supercritical accretion
regime \citep{ss2004}. Although its X-ray luminosity is much lower than
the expected Eddington luminosity ($\sim 10^{36}\div 10^{37}\ergl$,
depending on the interstellar absorption value), similar
objects observed at low inclinations $\lesssim 20^\circ$ are predicted
to be objects of high ($\gtrsim 10^{39}\ergl$) X-ray luminosity
\citep{katz86, poutanen}.
It allows to link \astrobj{SS433} and the still rather dim concept of supercritical
accretion with extragalactic Ultraluminous
X-ray sources (ULXs) known to have X-ray luminosities $10^{39}\div
10^{40}\ergl$. Though not unique in this sense, the supercritical
accretion model qualitatively
explains most of the observational properties of ULXs,
including their environmental and statistical properties
\citep{swartz04,list}. 

This interpretation is indirectly supported by the fact that some ULXs are
identified with powerful nebulae \citep{PaMir} with sizes from about
10 to hundreds of parsecs. The optical properties of
these nebulae are very diverse (see \citet{list} for review). Some
appear large-scale shock-powered bubbles. However, in several cases
an Extreme Ultraviolet (EUV) ionisation source is needed to
explain the properties of the nebula or some part of it or some
particular emission lines. In the case of the 
large nebular complex \astrobj{MH9/10/11}
around \astrobj{HoIX~X-1} \citep{hoixmois}, two regions powered by shocks
 and EUV radiation are clearly distinguished by kinematics and emission-line
 spectrum. 

 \astrobj{SS433} is surrounded by the radionebula \astrobj{W50} \citep{w50_dubner98} known
 to harbour optical filaments
\citep{zealey,w50_vdbergh}. Recent spectroscopic results
\citep{boumis,w50ifp} show that bright \nii\ and \oiii\ 
emissions are broadened by
$\sim 50\,\kms$. Their relatively high intensities (in comparison with Balmer
emissions) are easy to explain if
Galactic abundance gradients are taken into account.
The structure of the nebula is seen in details, but it is
difficult to infer its integral properties such as total line
luminosities and line ratios because of heavy and patchy absorption
toward the nebula and its large angular size (about $1\deg$). 
 The total emission line spectrum of W50 may
differ significantly from the spectra of the bright filaments. 

There were several works concerning the evolution of W50
\citep{zavala,velazquez}. They agree about the leading role of the
relativistic jets of \astrobj{SS433} in forming the peculiar ``seashell'' shape
of the nebula, but its central round core is usually attributed to the
pre-existing supernova remnant that is an essential element of all the
models. It is however unclear whether the supernova cavity is
required to reproduce the observed morphology. Another complication is
the non-relativistic wind of the supercritical disc of \astrobj{SS433}
\citep{ss2004}. Its power is about an order of magnitude less than
that of jets, but still sufficient to provide the
energy of an average supernova explosion of 
$\sim 10^{51}{\rm erg}$ in $\sim 10^5\rm yr$
period. Though it is natural to
expect the compact object in \astrobj{SS433} to be formed in a supernova
explosion, the significance of its
contribution to the power of the nebula is uncertain.

Here we make some estimates of the properties that a nebula produced
by an SS433-like object should have when 
evolving in a constant-density environment. 
We consider two main energy sources: mildly relativistic jets with
the kinetic luminosity of $L_j = 2\times 10^{39}\ergl$ and a photoionizing
anisotropic EUV/X-ray radiation source with a polar-angle-dependent
SED calculated according to \citet{AKK}.
There are also energy sources like the supernova explosion that was
likely to precede the formation of the compact accretor and a slow wind similar
to that of \astrobj{SS433} \citep{ss2004}. Both processes are expected to be
much less energetic than the jet activity and photoionizing radiation
of the object.

In the next section we consider the basic properties of the 
HII-region produced by the 
photoionizing radiation of the
central source. 
 In section~\ref{sec:photoin} we present several quasi-2D
photoionization models. In section~\ref{sec:bubble} we consider the
evolution of a jet-blown (beambag) nebula. We compare the results with
the existing properties of ULXNe and related objects (including W50
and the putative relic ULX bubble in IC10) and discuss our results in
section~\ref{sec:disc}.

\section{Str\"omgren Zone}\label{sec:hii}

Supercritical accretors are expected to be anisotropic sources due to
geometrical and relativistic 
beaming effects. Both X-ray radiation and some part of the
EUV photons (that are most relevant for nebular physics) escape only
in the certain
range of polar angles forming a complex non-spherical HII-region. 
We introduce here a generalized version of the Str\"omgren radius:

\begin{equation}\label{E:str}
R_S = \left( \frac{s}{\alpha n_e^2} \right)^{1/3}
\simeq 260\ n_0^{-2/3} s_{50}^{1/3} \pc
\end{equation}
\noindent

\begin{figure}
 \centering
\includegraphics[width=0.8\textwidth]{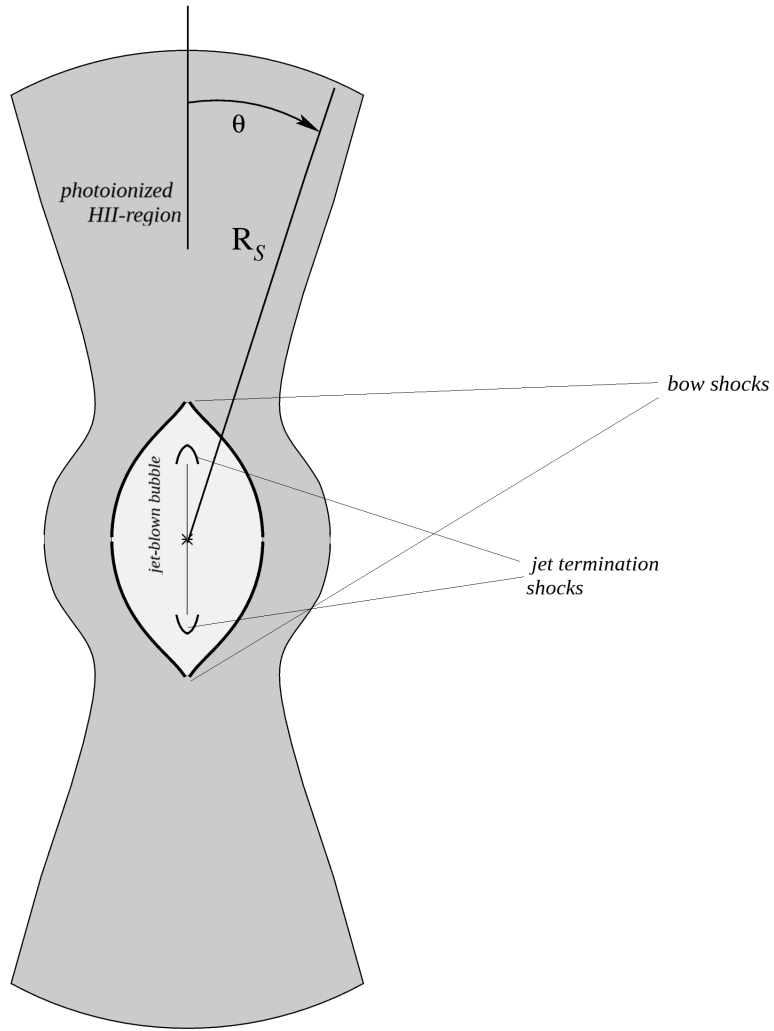}
\caption{Principal scheme of a hybrid nebula produced by a
  supercritical accretor. Protruding parts of the nebula are produced
  by the anisotropic radiation field.
}
\label{fig:nebscheme}
\end{figure}

 Here $n_0$ is hydrogen density, $\alpha$ is hydrogen recombination
 coefficient (see for example \citet{osterbrock}), $s = dS/d\Omega$ is
 the number of hydrogen-ionizing quanta
emitted in a unit solid angle. $s_{50}$ is $s$ in $10^{50}\,\rm s^{-1}$
units. For an isotropic source, $R_s$ is simply the radius of its Str\"omgren
sphere. A strongly anisotropic source produces a bilobial or elongated
nebula like the one shown in figure \ref{fig:nebscheme}.

A quasi-2D sectoral approximation is more than sufficient for
order-of-magnitude estimates. 
Deviations from this approximation are mainly due
to re-emission of absorbed ionizing quanta and are therefore $\sim
(\alpha_A-\alpha_B) / \alpha_A \sim 0.1$, where $\alpha_{A,B}$ are
recombination coefficients for the A and B cases, correspondingly
\citep{osterbrock}.
Deviations from the sectoral approximation are expected to be of the
order 10\%. 
 As long as these effects are not very important
one may consider the nebula extended by $R_S(\theta)$ in any direction
characterised by polar angle $\theta$. In the next section, we
calculate the structure of the nebula in more detail but still in
a sectoral approximation: at any polar angle $\theta$, the radial structure
is calculated as for a spherical nebula using our supercritical
accretion disc model spectrum for appropriate inclination angle $i=\theta$.

A Str\"omgren zone has a characteristic expansion time equal to the
recombination time \citep{DoSutISM}:
\begin{equation}\label{E:strt}
t_S = 1/\alpha n_e \simeq 1.2\times 10^5 T_4^{0.85} n_0^{-1} \yr
\end{equation}
\noindent
The expansion proceeds roughly according to the law $R(t) \propto
\left(1-\exp\left\{-t/t_s\right\} \right)^{1/3}$. Internal pressure
leads to further expansion of the HII-region but the expansion
velocities are close to the thermal and are therefore overridden by the
expanding jet-blown bubbles. We discuss this in more detail in
section~\ref{sec:bubble}.

We use our model of an outflow-dominated supercritical disc \citep{AKK}
to calculate the inclination-dependent
spectral energy distribution. It is dominated by X-rays and extreme
ultraviolet (EUV) radiation inside
the funnel (for polar angles less than the funnel half-opening angle,
$\theta_f \sim 20^\circ$)
and by UV and optical at larger angles. Extended low-inclination
parts of the nebula receive a broad-band ionizing spectrum rich in
soft X-rays and in EUV, similar to that of active galactic nuclei (AGN).

X-ray ionized nebulae (XINe) differ from ordinary nebulae powered by
EUV radiation in several ways. One is that XINe do not
have any sharp outer boundary but the boundary is determined by either
absorption of X-rays or the equilibrium temperature.  
The outer radius of a XIN may be estimated as the radius of
penetration of standard or soft X-rays bearing most of the energy. 

\begin{equation}\label{E:xin}
R_X \simeq \frac{1}{n \sigma(E)},
\end{equation}
\noindent
where $\sigma(E)$ is the absorption cross-section at energy $E \sim
T \sim 1\,\keV$ below which most of the quanta are emitted. 
Using the simple approximation of \citet{haya}, one
arrives to the following estimate:

\begin{equation}\label{E:xin1}
R_X \simeq 1.6 T_1^{2.5} n_0^{-1} \kpc
\end{equation}
\noindent

$T_1$ here is the characteristic temperature of the energy source in \keV.
The electron temperature fades away in XINe in a power-law fashion
(roughly as $T
\propto r^{-1/2}$, see \citet{xine_rappaport}) therefore the
definition of the outer radius of a nebula powered by X-rays is
largely uncertain. 

\section{Quasi-2D Modelling}\label{sec:photoin}

In order to understand the basic properties of the photoionized
nebulae we suggested that the Str\"omgren zone is established in a
quasi-2D way: the structure of the HII-region in every given
direction is determined by the generalized quanta production rate only
(that is a function of the polar angle). For every polar angle value,
a spherical model was calculated. A set of 50 spherical models for
polar angles between $0$ and $\pi/2$ was calculated. 
Three series of spherical {\it Cloudy} (version 08.00) models were computed: one
for a low ISM density (1$\cmc$) and solar metallicity, one for an
HII-region density of 100$\cmc$ and solar metallicity, and one
for the high density of 100$\cmc$ and metallicity ten times less then solar.
The inner radius is set to $10\pc$ in order to reproduce the compact inner
cavern created by shocks. The input spectrum was calculated using the
supercritical funnel model by \citet{AKK} with $\mdot=100$,
$\theta_f=0.4{\rm rad }\simeq 23^\circ$, $r_{in} = 1/6\mdot = 1/600$ and self-irradiation
accounted for by two iterations (see original work by \citet{AKK} for
details). 

\begin{figure*}
 \centering
\includegraphics[width=\textwidth]{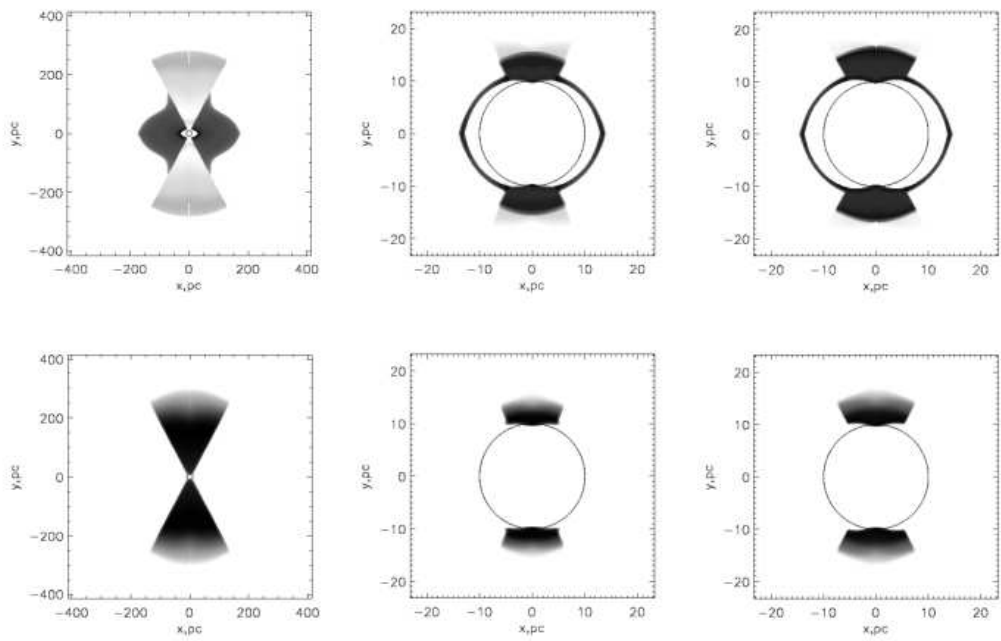}
\caption{
Model emissivities of recombination lines: H$\beta$ (upper row) and
\heii$\lambda$4686 (lower row). From left to right the pictures
correspond to the low-density, high-density and low-metallicity
models. 
}
\label{fig:reclines}
\end{figure*}

\begin{figure*}
 \centering
\includegraphics[width=\textwidth]{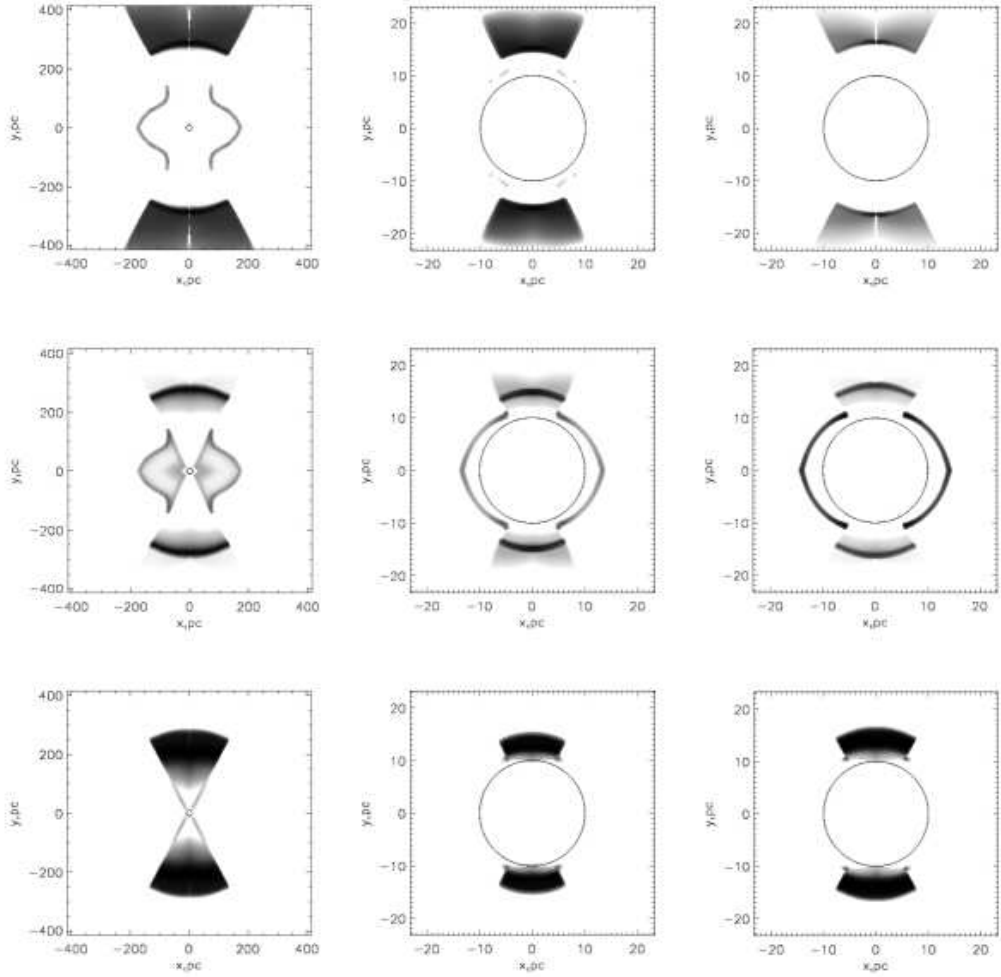}
\caption{
Model emissivities of oxygen lines excited by collisions (from top to bottom): 
\oi$\lambda$6300, \oii$\lambda$3727 and \oiii$\lambda$5007. The models
are in the same order.
}
\label{fig:collines}
\end{figure*}

\begin{figure*}
 \centering
\includegraphics[width=\textwidth]{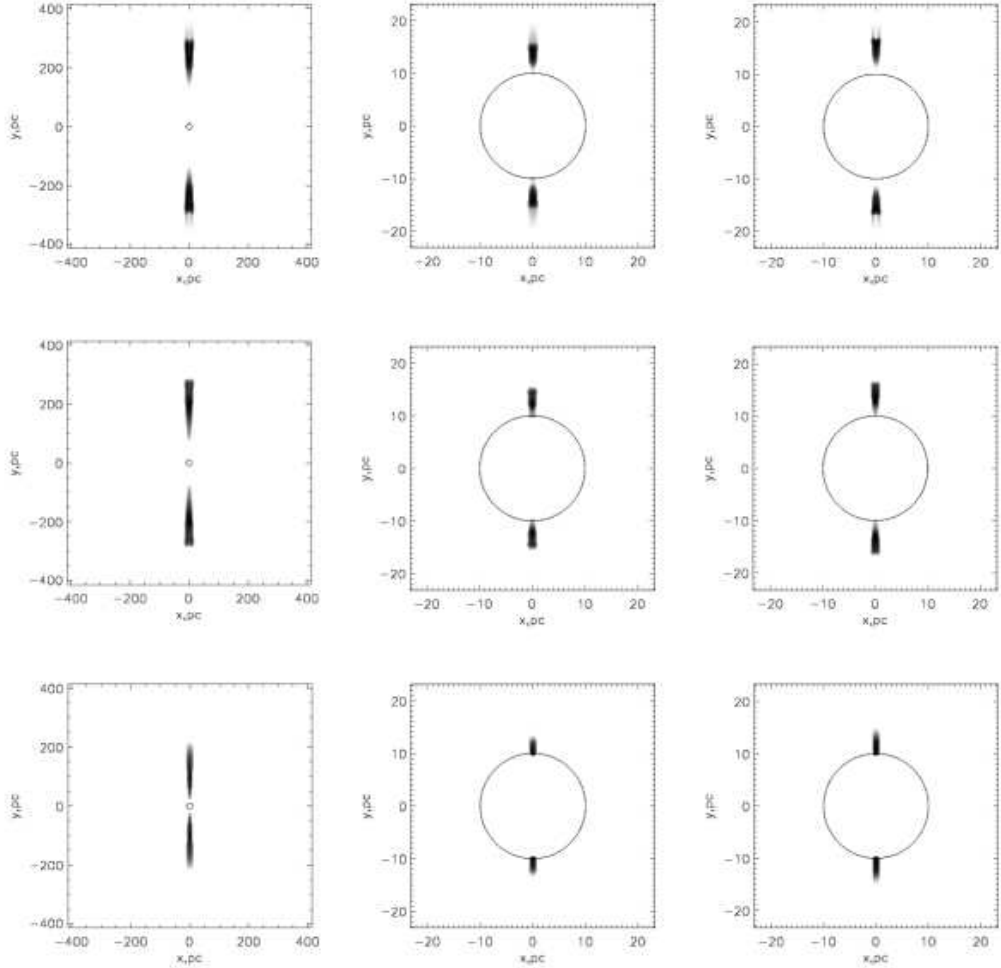}
\caption{
Model emissivities of emission lines sensitive to heating: \hei$\lambda$4471 (upper row),
\oiii$\lambda$4363 (second row)
and ``coronal'' \fevii$\lambda$6078 (bottom). The models
are in the same order.
}
\label{fig:hotlines}
\end{figure*}

In figures \ref{fig:reclines}$\div$\ref{fig:hotlines}, we show the
two-dimensional emissivity maps
(cross-sections) of the model
nebulae using combined sets of {\it Cloudy} simulations. The scale
everywhere is linear.
As may be seen in the figures, a nebula usually has two
characteristic radii, one corresponding to the Str\"omgren zone of the
disc wind photosphere and the other to the nebula produced by the funnel
interior. This structure is well established in recombination lines
(see figure \ref{fig:reclines}). The emissivity of a collisionally
excited line is high in a narrow layer
where the relevant ionic species is abundant and fades smoothly away outside the
Str\"omgren radius where ionization by X-rays is important (see figure \ref{fig:collines}). 
Other lines such as \hei$\lambda$4471
and coronal emissions are sensitive to additional heating (see figure \ref{fig:hotlines}).

\begin{table*}
\caption{Integral luminosities ($10^{37}\ergl$ units) of some selected lines for the three
  {\it CLOUDY} models described in section
  \ref{sec:photoin}. }\label{tab:lines} 
\center{
\begin{tabular}{l||p{3.5cm}|p{3.5cm}|p{3.5cm}}
line ID & 1\cmc, solar metallicity & 100\cmc, solar metallicity &
100\cmc, 0.1 solar metallicity \\
\hline
\noalign{\smallskip}
H$\beta$  & 100  & 106 & 110 \\
\oiii$\lambda$5007 & 990 & 1260 & 320 \\
\oii$\lambda$3727 & 270 & 154 & 60 \\
\oi$\lambda$6300 &  44 & 39 & 13 \\
\heii$\lambda$4686 & 38 & 21 & 20 \\
\sii$\lambda$6717 & 28 & 30 & 13\\
\fevii$\lambda$6087 & 5.1$\times 10^{-3}$ & 8.3$\times 10^{-3}$& 1.8$\times 10^{-3}$\\
\hei$\lambda$4471 & 8.7$\times 10^{-3}$& 7.1$\times 10^{-3}$ & 7.4$\times 10^{-3}$ \\
\oiii$\lambda$4363 & 0.044 & 0.066 & 0.037 \\
\feiii$\lambda$4659 & 4.3$\times 10^{-3}$ & 2.2$\times 10^{-3}$ & 4.2$\times 10^{-4}$\\
\end{tabular}
}
\end{table*}

The luminosities of selected emission lines calculated with the three models are
given in table~\ref{tab:lines}. Note the HeII$\lambda$4686 to H$\beta$
flux ratios about 0.2$\div$0.3 and very high [OIII]$\lambda$5007 to
H$\beta$ flux ratios around 10 (for higher metallicity). These values together with rather
bright collisionally-excited lines of low ionization potentials like
\oi$\lambda$6300 and \sii$\lambda$6717 are consistent with the
spectra of ULX nebulae proposed as HII-regions excited by the hard
central radiation source such as the nebula of M101X98 \citep{list}
and MH11 \citep{hoixmois}.
 The properties of some individual ULX
nebulae will be discussed in section~\ref{sec:disc}. 
It should be also noted that the real nebular spectra are usually
obtained with either a slit or a finite-size aperture that distorts
the observed spectrum. Luminosities of low-ionization emissions may be underestimated
by observational data. 

\section{Bubble Nebulae}\label{sec:bubble}

\subsection{Jet Propagation Law}

The expected evolution of the nebula produced by relativistic jets is
similar to the scenario proposed by \citet{BC89} for AGN. Unlike the
intergalactic and galactic media in the case of radiogalaxies,
the interstellar medium density is more or less constant that prevents
ULXs from forming extended shock-powered lobes. 

The jet propagation is governed by the momentum balance at the jet head. In
the reference system of the bow shock, the ram pressure of the
unshocked material plus thermal pressure of the medium 
should be balanced by the momentum flux in the jet equal to $\beta
(\gamma -1 ) L_j / 2 c \Omega_j $ where $\Omega_j$ is the solid angle
of a single jet. If jets are precessing or jittering,  $\Omega_j$
effectively will be larger. $L_j$ is the luminosity of the jet pair.

We neglect the pressure of the unshocked jet material. In the case of
\astrobj{SS433}, the gas pressure is about eight orders of
magnitude less than the bulk-motion momentum flux in the region where
moving emission lines are formed. If this ratio holds,
contribution of the jet gas pressure is always negligible. This
statement is even stronger if the gas is a subject to radiative or
adiabatic cooling.

\begin{equation}\label{E:jetbalance}
\rho \left(\frac{dz}{dt}\right)^2 + P = \frac{\beta ( \gamma - 1 )
  L_j}{2 c \Omega_j z^2}
\end{equation}
\noindent
The solution may be found analytically. There is a characteristic time $t_j$ of
jet deceleration, and a maximal length of the jet $R_j$ that can not be
exceeded. Until $R_j$ is reached, jet head moves according to the law:

\begin{equation}\label{E:jetlaw}
\zeta = \sqrt{\tau \left( 2 - \tau \right)},
\end{equation}
\noindent
where $\zeta$ and $\tau$ are correspondingly the dimensionless 
jet length and age, $\zeta = z / R_j$ and $\tau = t / t_j$.
Final position of the jet head in this approximation equals:

\begin{equation}\label{E:jetR}
R_j \simeq 36 \sqrt{\frac{\Omega_j}{\gamma_j \beta_j}} L_{39}^{1/2}
n_0^{-1/2} T_4^{-1/2}\ \pc
\end{equation}
\noindent

Here, $L_{39}$, $n_0$ and $T_4$ are the jet kinetic luminosity in
$10^{39}\ergl$, interstellar medium hydrogen density in \cmc\ and its
temperature in $10^4\,K$ (note that the gas is ionized and
heated by the radiation of the central source). 
The jet head position is stationary for $t>t_j$
because the interstellar pressure $P$ becomes significant. 

Characteristic time scale $t_j$ of jet deceleration equals:

\begin{equation}\label{E:jett}
t_j \simeq 3 \sqrt{\frac{\Omega_j}{\gamma_j \beta_j}} L_{39}^{1/2}
n_0^{-1/2} T_4\ \Myr
\end{equation}
\noindent
\bigskip 

The jet bubble size is the size of a pressure-driven bubble filled with
hot gas and expanding according to the well-known law established by
\citet{avedisova} (see also \citet{castor} and \citet{lozinsk}):

\begin{equation}\label{E:R:JB}
R_{JB} \simeq 100 L_{39}^{1/5} n_0^{-1/5} t_6^{3/5} \pc,
\end{equation}
\noindent

where $t_6$ is time in million years (Myr). Radius of a wind-blown
bubble may
be estimated in the same way. In the case of \astrobj{SS433}, the
power of the isotropic wind is about one
order of magnitude lower.

\begin{equation}\label{E:R:WB}
R_{WB} \simeq 70 L_{w,38}^{1/5} n_0^{-1/5} t_6^{3/5} \pc
\end{equation}
\noindent

The characteristic radii as functions of time for fixed ISM density 
and other parameters are shown in figure~\ref{fig:lums_radii}. 

\subsection{Line Luminosities}

Below we use the luminosities of Balmer lines as characteristic. Two
energy sources are expected to produce strong nebular lines: photoionization
by the central X-ray/EUV source and radiative shock waves. One may
estimate the contribution of the shock-powered part of the nebula as:

\begin{equation}\label{E:hbshock}
\begin{array}{l}
  L_S(H\beta) = \left( 3.26\times 10^{-3} \,V_2^{-0.59} +\right. \\
\qquad{} \left. +4.32\times 10^{-3} \, V_2^{-0.72} \Theta(V_2-1.5)
\right) L_j,\\
\end{array}
\end{equation}
\noindent
where $L_j$ is the mechanical luminosity of the wind or jets,
 $V_2$ is the velocity (in $100~\kms$ units) of the shock actually
producing the optical emission. $\Theta$ here is the Heaviside function,
equal to 0 if its argument is less than 0 and 1 otherwise. It is used
to switch on the shock wave precursor for velocities
higher that about $150~\kms$ \citep{DoSutI}. 

The $H\beta$ luminosity of the HII-region may be estimated as:

\begin{equation}\label{E:hbQ}
L_H(H\beta) \simeq 5\times 10^{37}  Q_{50} \ergl
\end{equation}
\noindent
$Q_{50}$ here is the total production rate of hydrogen-ionizing quanta
in $10^{50}\,\rm s^{-1}$ units. An EUV source with a luminosity $L_{EUV} \sim
10^{40}\ergl$ provides $Q \sim 4\times 10^{50}\,\rm s^{-1}$. 

Several important details should be mentioned concerning emission
line luminosities: {\it (i)} shock waves should become radiative in order to
produce optical nebular emission; {\it (ii)} shock emission
disappears if the velocity is very low, $V_S \lesssim 20\kms$; {\it
  (iii)} a shock wave leaves a cavern of rarefied gas
practically transparent to ionizing radiation. Generally,
photoionization produces brighter lines but the shape of the nebula is
determined by shock waves. 

We assume that the primary source of mechanical energy is a pair of jets with a total
power $L_j = 2\times 10^{39}\ergl$ in order to simulate an SS433-like
system.
Below we will trace the evolution of such a system for several
million years. All the jet power is supposed to be distributed over a
given solid angle $\Omega_j$ due to precession or jitter; we assume
$\Omega_j = 0.038\,\rm sr$.
 If one considers an SS433-like object with a
stationary jet ($\Omega_j \sim 10^{-3}\div 10^{-4}$), $R_j$ becomes
much higher, about $1\div 3\,\kpc$. As far as I am concerned, structures of this
kind (kiloparsec-scale double-lobe nebulae) have not ever been
observed. ISM density is likely to vary significantly on these spatial
scales. In a spiral galaxy, the most likely fate of a kiloparsec-scale
jet is leaving the galactic disc and producing a fountain.

Characteristic time scales are: the Str\"omgren zone establishment time, $t_S$,
the jet propagation time, $t_j$, the shock cooling (to temperatures when the
shocked gas becomes radiative; see for example \citet{castor}) time
$t_{cool}$.

\begin{equation}\label{E:stromt}
t_S \simeq  1 Q_{50}^{1/2} n_0^{-1/2} T_4^{-3/4}  \Myr
\end{equation}
\noindent

\begin{equation}\label{E:jett}
t_j \simeq 3 \sqrt{\frac{\Omega_j}{\gamma_j \beta_j}} L_{39}^{1/2}
n_0^{-1/2} T_4\ \Myr
\end{equation}
\noindent

\begin{equation}\label{E:coolt}
t_{cool} \simeq 3\times 10^4 L_{39}^{1/2} n_0^{-1/2} \ \yr
\end{equation}
\noindent

\begin{figure}
 \centering
\includegraphics[width=1.2\columnwidth]{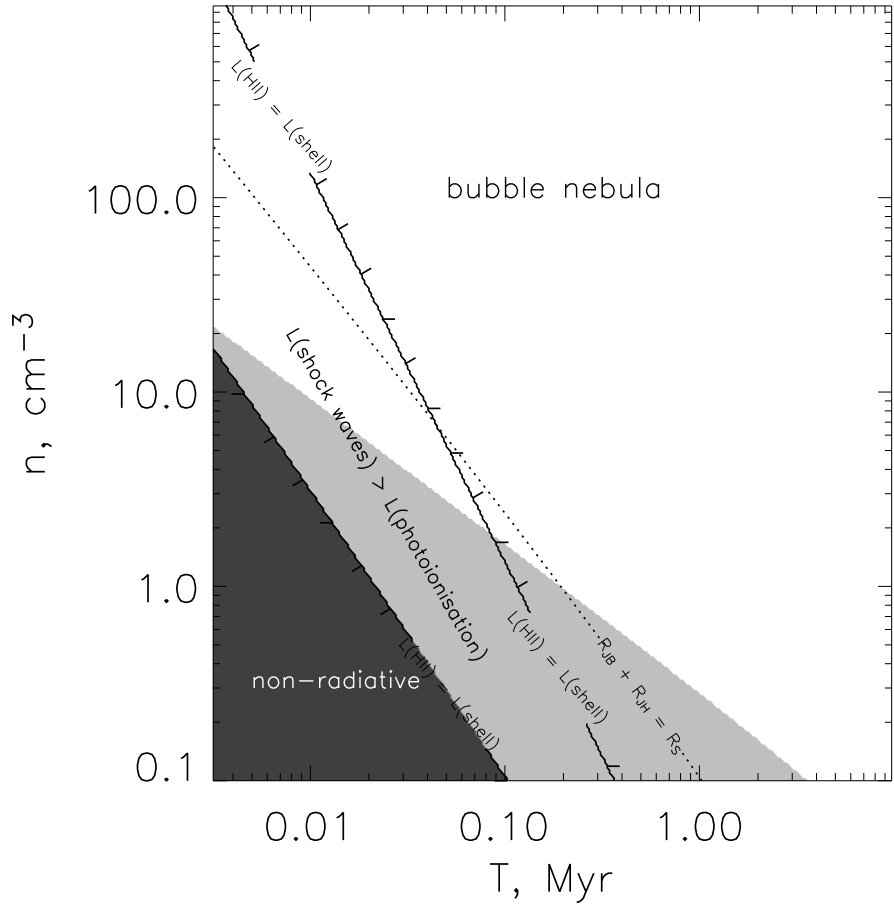}
\caption{
 Emission regimes of a nebula powered by a supercritical
 accretor. Shock waves are non-radiative in shaded
 regions (in the darkest area, both bow shocks and bubbles are
 non-radiative). The straight line with ticks demarcates the regions where
 photoionization and shock waves are expected to dominate the Balmer line
 emission, respectively. At the dotted line, the sizes of the jet-blown bubble and
 the Str\"omgren zone are approximately equal.
}
\label{fig:lums}
\end{figure}

\begin{figure}
 \centering
\includegraphics[width=1.2\columnwidth]{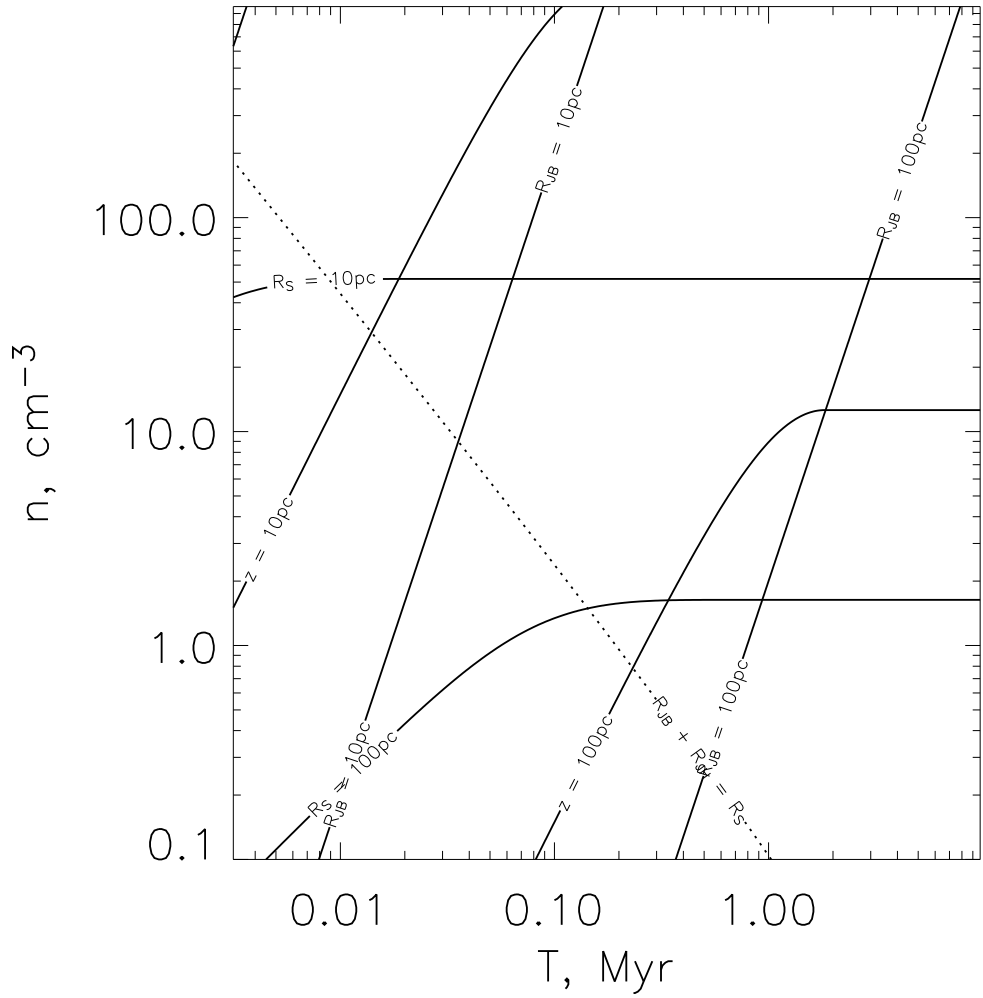}
\caption{
Various characteristic radii as functions of system age and ISM
density. $R_S$, $R_{JB}$ and $z$ are correspondingly the Str\"{o}mgren
radius, jet-blown bubble radius and jet length. See text for details.
}
\label{fig:lums_radii}
\end{figure}

\begin{figure}
 \centering
\includegraphics[width=1.2\columnwidth]{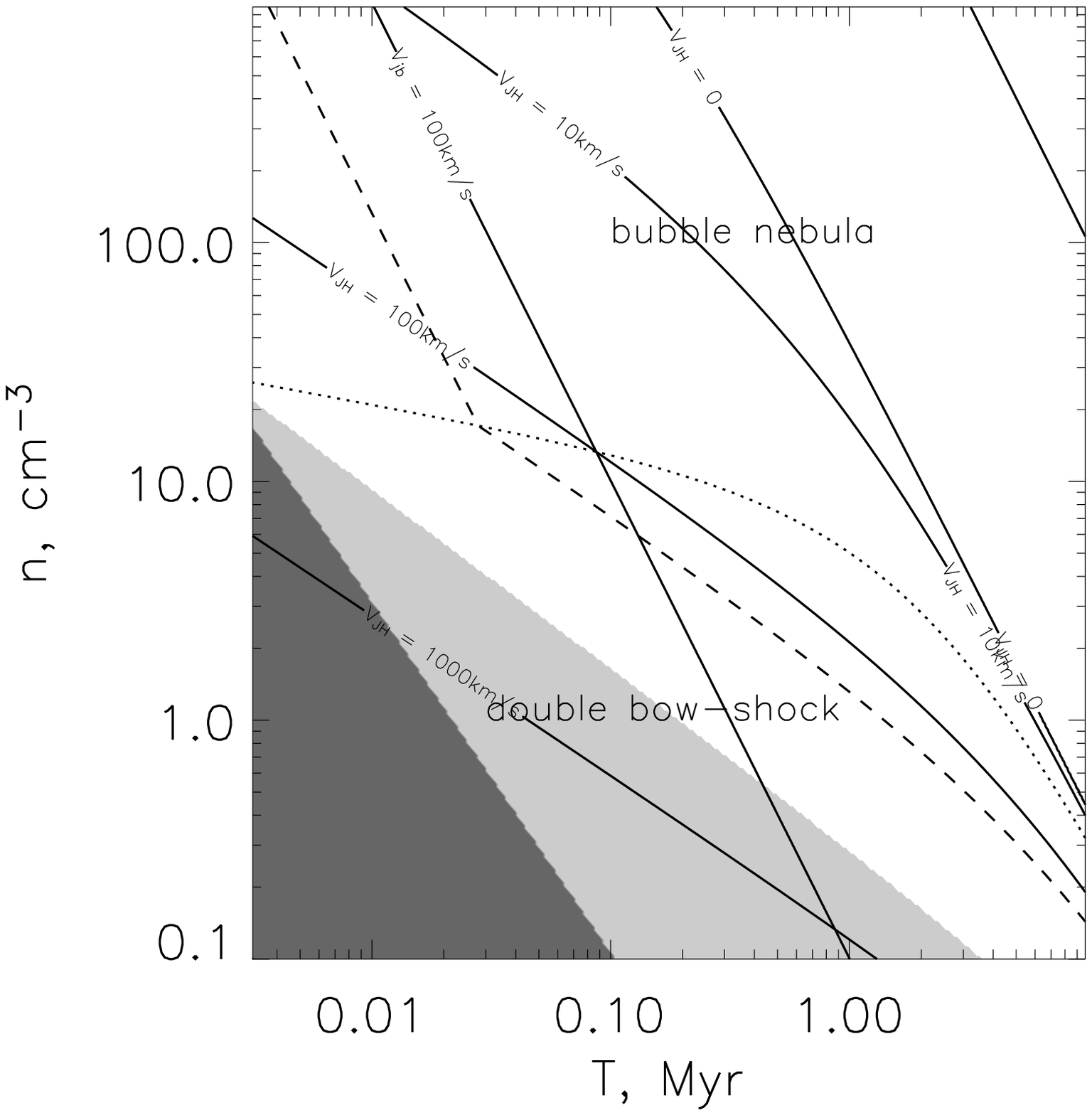}
\caption{
Jet head and jet bubble expansion velocities as functions of the
age and ISM density. The dotted line shows the boundary between the
double bow shock case (jet propagation velocity is higher than the bubble
expansion velocity) and the bubble/multibubble stage. The dashed line
demarcates the regions of fast (with precursors) and slow shock
waves. In the shaded region, shocks are non-radiative (cooling time is
comparable to or higher than the age of the system).
}
\label{fig:vels}
\end{figure}

In figures \ref{fig:lums}-\ref{fig:vels}, we illustrate the dependence of
various parameters 
of the nebula on time and ISM density.
Generally, the nebula around a supercritical accretor is powered by
photoionization until shock waves become radiative. 
Its properties are expected to be similar to the
model nebulae described in the previous section, but the hot cavern in
the center expands with time. 

A double bow shock nebula may exist even at longer times but usually jet
bubble expansion velocities are higher and the nebula appears to be
either a bubble or a multi-bubble shock-powered shell. Bubble
coalescence is a poorly constrained process. Under almost any
possible conditions, expansion of the two jet-powered bubbles is fast enough to produce a
single cocoon at the very moment of jet launching. 
 
\subsection{Double Bow Shock and Bubble Stages}

The HII-region is torn from inside by the expanding jets and jet
bubbles. I consider the nebula in the double bow shock dominated stage if
the jet head propagation velocity is higher than the jet bubble
expansion velocity. The bubble expansion is expected to be highly
anisotropic in the ``bow shock'' stage but the details are uncertain
and depend on the gas flows inside the bubble. I would also expect
the morphology of the shell to depend on the ratio of the two
velocities. I assume the nebula is in the ``double
bow shock'' stage if the jet propagation velocity is higher than bubble
expansion velocity, and in the ``single bubble'' stage otherwise. 

 Generally, ``double bow shock'' nebulae exist in
rarefied medium indicating that there
should be a correlation between the elongated or bilobial form of the
nebula and the low density of the ambient medium.
ULXs located inside star-forming regions or molecular clouds are
expected to evolve rapidly toward the quasi-isotropic single bubble
stage. 

\section{Discussion}\label{sec:disc}

\subsection{Comparison with Observations}

The large diversity of the observational properties of ULX nebulae may
be explained (at least qualitatively) in the framework of a hybrid model
taking into account both EUV radiation and shock waves. Let us see how
individual objects fit into this unified scheme. 

{\it \astrobj{MF16}} is a well-known but outstanding example. It is the most compact
of all the known ULXNe and has the highest expansion rate
measured ($V \sim 200\kms$). It was shown to have an elongated
bilobial morphology with a fainter extended halo surrounding a
prominent shell with a radius of $10\div 20\pc$ \citep{BFS}.
All these parameters are consistent with the position around $n_0 \sim
5\,\cmc$, $t \sim 30\,000\,\rm yr$ in our diagrams. The extended halo probably
corresponds to the outer part of the HII-region that is expected to be
larger than the jet bubble radius at this age. The ISM density was
estimated in \citet{mf16_pasj} and equals $n_0 \simeq 6\,\cmc$.
The uniqueness of MF16 is probably the result of its evolutionary stage that
lasts less then one tenth of the expected lifetime of the source. 

Several objects selected by \citet{PaMir} are large-scale bubbles with
dynamical ages $\sim 1\Myr$ and quasi-spherical morphology probably
affected by underlying density gradients. It is likely that their
expansion started in a rather dense environment of the host
star-forming region that implies an effective density value of $n_0 \sim
10\div 100\,\cmc$. 

Predicted lifetimes for nuclear-timescale accretion in massive X-ray
binaries are of the order 1\Myr\ \citep{RPP}, which is comparable to
the dynamical ages of ULX bubbles. It is possible that the
bubble survives the further evolution of the object that is expected to
evolve toward a binary system similar to \astrobj{Cyg~X-3} or \astrobj{IC10~X-1} \citep{ic10}, consisting of
a black hole and a Wolf-Rayet donor. In this scope, the giant
synchrotron superbubble around \astrobj{IC10~X-1} \citep{lomo} is the best
candidate for a relic ULX bubble. Its kinematical properties are close
to those of the largest ULX bubble nebulae \citep{funas}. The concurrent explanation of the
bubble in IC10 as a hypernova remnant can not account for this
similarity. 

{\it \astrobj{HoII~X-1}} does not show any signatures of supersonic motions or
contamination with heavy elements \citep{lehmann}. The nebula appears
to be larger then the Str\"omgren radius inferred from recombination
line luminosities \citep{hoixmois,mf16_pasj} that may be 
explained by a
non-radiative central cavern created by the preceding supernova
explosion and/or wind/jet activity.
This is consistent with the observed non-thermal radiospectrum of the
nebula. 

{\it \astrobj{W50}} is also quite easy to place on the diagram. The undisturbed ISM
density $\sim 2\,\cmc$ is known from \citet{lockman_w50} so the
nebula may be considered a prototype of an evolved ULXN in a rarefied
environment. Its properties are consistent with an age about several
$\times 10^5\,\rm yr$ and ISM density  $\sim 2\,\cmc$. 
These estimates imply that the bubble expansion velocity for W50 should be
$V_{JB} \sim 50\div 100\kms$ consistent with velocity estimates by
\citet{zealey, boumis, w50ifp}. 

\subsection{Jet-Powered Nebulae}

The ambiguity of the energy sources of ULXNe bears some similarity with
the question about the power sources of Seyfert
Narrow-Line Regions (NLRs). This similarity is most striking for MF16
\citep{mf16_garching} where the integral spectrum itself is similar to
that of a Seyfert NLR. 

Another parallel with AGN is in the pair of relativistic jets that in
the case of SS433-like objects are likely to produce something similar
to the cocoons that were predicted \citep{BC89} but 
are not observed around AGN mainly due to strong
gradients in ISM and IGM density. 

Little can be said for certain about the structure of a nebula
produced by relativistic jets. However, W50, ULXNe and similar objects
are prospective targets for sophisticated hydrodynamical modeling. For
example, in the works by \citet{velazquez} and \citet{zavala},
hydrodynamical simulations allowed to qualitatively explain the basic
properties of W50. I suggest that running simulations at slightly
different parameters would allow 
to better understand the diversity of ULXNe that
sometimes evolve into quasi-spherical bubbles, but under different
conditions may retain a
resemblance to W50 instead (as in the case of MF16).

\section{Conclusions}

In this article, I have shown that most of the observational properties of
ULX nebulae may be explained by a single unified scheme. Only two
free parameters were varied (ISM density and the age of the source), 
though some divergence in the parameters of the
binary system itself is expected. 

One of the main results is the bipolar morphology 
in some ``hot'' lines such as \heii$\lambda$4686 and \oiii\
emissions. The ``ionization cone'' structures detected by \citet{rgww} and
the elongated high-excitation nebulae around M101P098 and
HoIX~X-1 are naturally explained in the framework of our model.
Bilobial and elongated shock-powered nebulae are expected to appear in
a low-density environment due to the different dependences of the jet
propagation and bubble expansion velocities on the ambient
density. The morphology of ULX shells is predicted to evolve through
a ``double bow shock'' bipolar structure toward a
quasi-isotropic single bubble stage.

Supercritical accretors are likely to exist for about 1\Myr (of the
order of the nuclear timescale of the donor star). 
 Then the
source is expected to evolve into a WR+BH binary, possibly
surrounded by a relic ULX bubble. We propose that the synchrotron
bubble in IC10 is a relic ULX bubble rather than a hypernova remnant.


\bibliographystyle{model2-names}
\bibliography{mybib}







\end{document}